\documentclass[epj,referee]{svjour}
\usepackage{graphics}

\begin{document}
\title{Acceleration of ultra-thin electron layer. Analytical treatment compared with 1D-PIC simulation.}
\author{Meng Wen\inst{1,2}, Hui-Chun Wu\inst{1}, J\"{u}rgen Meyer-ter-Vehn\inst{1}, \and Baifei Shen\inst{2}}
\institute{Max-Planck-Institut f\"{u}r Quantenoptik, D-85748 Garching, Germany
\and Shanghai Institute of Optics and Fine Mechanics, Shanghai 201800, China}
\date{Received: date / Revised version:}

\abstract{
In this paper, we apply an analytical model [V.V. Kulagin et al., Phys. Plasmas 14,113101 (2007)] to describe the acceleration of an ultra-thin electron layer by a schematic single-cycle laser pulse and compare with one-dimensional particle-in-cell (1D-PIC) simulations.
This is in the context of creating  a relativistic mirror for coherent backscattering and supplements two related papers in this EPJD volume.  The model is shown to reproduce the 1D-PIC results almost quantitatively for the short time of a few laser periods sufficient for the backscattering of ultra-short probe pulses.
  }
\authorrunning{M. Wen, H.-C. Wu, J. Meyer-ter-Vehn, B. Shen}
\titlerunning{Acceleration of ultra-thin electron layer}
\maketitle
\section{Introduction}
Irradiation of ultra-thin solid foils by high-contrast laser pulses
at relativistic intensities may provide a new way to create a
relativistic mirror for coherent reflection, generation of high
harmonics and compression of drive and probe light
\cite{Kulagin,sliding mirror,Vshivkov,Shen,Mikhailova}. We consider
the regime in which foil electrons are completely separated from
ions \cite{paper I}. The partial reflectivity of such electron
layers has been derived and compared with one-dimensional
particle-in-cell simulation in \cite{paper II}. Coherent reflection
depends on the square of the layer density and is therefore
sensitive to the density profile. The analytical model developed by
Kulagin et al. \cite{Kulagin} allows to describe the layer evolution
almost quantitatively, at least for a few laser periods after
expulsion from the foil. This time interval is sufficient to reflect
few-cycle probe pulses, to compress them to atto- and zepto-second
duration, and to shift their spectra to the VUV- and X-ray regime.
Here we apply the Kulagin model to a reference case used in the
previous publications in order to develop and explore this method.

\section{Analytical Model}

Different from \cite{paper I}, we consider here the electromagnetic field
$\mathbf{E}=\mathbf{E_L}+\mathbf{E_s}$ and $\mathbf{B}=\mathbf{B_L}+\mathbf{B_s}$
including the self-fields $\mathbf{E_s}$ and $\mathbf{B_s}$ of the electron
layer in addition to the external plane laser field, propagating in $x$-direction
and having linear polarization with the components $E_{Ly}$ and $B_{Lz}$.
Throughout this paper, we use natural units, i.e. time and space coordinates
are normalized according to $t'=\omega_L t$, $x'=k_L x$, where $\omega_L$
$k_L$ are laser frequency and wave number, fields $E'=eE/(mc\omega_L)$,
$B'=eB/(mc\omega_L)$, velocities $\beta=v/c$, and momenta $p'=p/mc$,
where $e$ is charge unit, $m$ electron mass, and $c$ velocity of light.
In the following, these normalized quantities are used dropping the prime.

The self-fields consist of the longitudinal electrostatic field $E_{sx}$
due to charge separation between electrons and ions and the transverse
electromagnetic fields $E_{sy}$ and $B_{sz}$ due to induced electron
currents. The laser pulse initially hits a foil which has uniform
electron and ion density $n_e=n_i=n_0$ and thickness $d_0$. While the ions
are taken as immobile heavy particles, the electrons move in the fields.
Their initial position $x_0$ in the layer ($0\le x_0 \le d_0$) is taken
as Lagrangian coordinate, and the goal is to determine the trajectories
$x(t,x_0)$ and $y(t,x_0)$.

The longitudinal field, felt by electrons of initial position $x_0$ when at
position $x$, is then given by
\begin{equation}\label{Esx}
E_{sx}(x,x_0) = \left\{ {\begin{array}{*{20}c}
   {N(x-x_0)} \quad & {(x_0\le x \le d_0)} \hfill  \\
   {N(d_0-x_0)} \quad & {(x \ge d_0) } \hfill  \\
\end{array}} \right.,
\end{equation}
where
\begin{equation}
N=n_0/n_{crit}=\omega_p^2/\omega_L^2
\end{equation}
is the layer density normalized to the critical density $n_{crit}$;
it can be conveniently expressed by the squared plasma frequency
$\omega_p^2=4\pi e^2 n_0/m$ divided by $\omega_L^2$. Equation (\ref{Esx})
describes the increase of $E_{sx}$ linear in $x$ for electrons still
inside the ion volume and the constant $E_{sx}$ after leaving
the ion layer. It is valid as long as electron sub-layers
with different $x_0$ keep their relative order.

Next we consider the electromagnetic fields due to the currents
induced in the electron layer by the driving laser. An electron,
that originated from $x_0$ and has position $x(t,x_0)$ and
velocities $\beta_x(t,x_0)$, $\beta_y(t,x_0)$ at time $t$,
experiences approximately
\begin{equation}
E_{sy}(t,x_0)  =\frac{N\beta_y}{2}\left[
\frac{x_0}{1-\beta_x}+\frac{d_0-x_0}{1+\beta_x}
\right] ,
\label{Esy}
\end{equation}
\begin{equation}
B_{sy}(t,x_0)  =\frac{N\beta_y}{2}\left[
\frac{x_0}{1-\beta_x}-\frac{d_0-x_0}{1+\beta_x}
\right].
\label{Bsz}
\end{equation}
Here the first terms in the square brackets stem from electron currents
to the left ($x_0'<x_0$) and the second terms from those to the right
($x_0'>x_0$) of the considered electron at $x_0$. The approximation
made in the present model is that the velocities $\beta_x$ and $\beta_y$
are assumed spatially uniform.

The equations of motion then are
\begin{eqnarray}\label{eom}
\frac{dp_{x}}{dt} &=&-E_{x}-\beta _{y}B_{z},  \nonumber \\
\frac{dp_{y}}{dt} &=&-E_{y}+\beta _{x}B_{z}, \nonumber\\
\gamma ^{2} &=&1+p_{x}^{2}+p_{y}^{2},  \nonumber \\
\frac{d\gamma }{dt} &=&-\beta _{x}E_{x}-\beta _{y}E_{y}, \\
\frac{dx}{dt} &=&\beta _{x}=p_{x}/\gamma , \nonumber\\
\frac{dy}{dt} &=&\beta _{y}=p_{y}/\gamma \nonumber.
\end{eqnarray}
From this we find
\begin{eqnarray}
&&\frac{d}{dt}(\gamma-p_x) =(1-\beta_x)E_{x}-\beta_{y}(E_y-B_z),  \nonumber \\
&&\frac{dp_{y}}{dt} =-(1-\beta_x)E_{y}-\beta_{x}(E_y-B_z).
\end{eqnarray}
For a plane wave moving in vacuum in $x$-direction,
the laser fields satisfy the dispersion relation $\omega_L=ck_L$ and
$E_{Ly}(\tau )=B_{Lz}(\tau )$ with propagation coordinate $\tau =t-x$.
For an electron moving along $x(t)$, this implies $d\tau /dt=1-\beta _{x}$.
Introducing $\kappa=\gamma-p_x$, making use of
$\kappa=\gamma(1-\beta_x)=\gamma d\tau/dt$,
and recalling $E_x=E_{sx}$, $E_y=E_{Ly}+E_{sy}$, $B_z=B_{Lz}+B_{sz}$,
we derive, after some algebra, the coupled equations
for $\kappa(\tau)$ and $p_y(\tau)$:
\begin{eqnarray}
\frac{d\kappa}{d\tau}&=&E_{sx} - N(d_0-x_0)\frac{p_y^2}{1+p_y^2},\label{kappa-eq} \\
\frac{dp_{y}}{d\tau } &=& -E_{Ly} - \frac{Nd_0}{2}\frac{p_y}{\kappa},\label{py-eq}.
\end{eqnarray}
They are solved by numerical integration.
From $\kappa=\gamma-p_x$, $\gamma^{2}=1+p_x^2+p_y^2$, and the solutions
$p_y(\tau)$ and $\kappa(\tau)$, one can obtain
\begin{eqnarray}
\gamma (\tau ) &=&\frac{1+p_{y}^{2}}{2\kappa }+\frac{\kappa }{2}, \label{gamma-eq}\\
p_{x}(\tau ) &=&\frac{1+p_{y}^{2}}{2\kappa }-\frac{\kappa }{2},
\label{px-eq}
\end{eqnarray}
and the particle trajectories then follow from
\begin{eqnarray}
dx/d\tau &=&p_{x}/\kappa , \label{x-eq}\\
dy/d\tau &=&p_{y}/\kappa ,\label{y-eq}
\end{eqnarray}
in parametric form with time given by $t(\tau )=\tau +x(\tau )$ .

\begin{figure}[b]
\centering\resizebox{1\columnwidth}{!}{\includegraphics{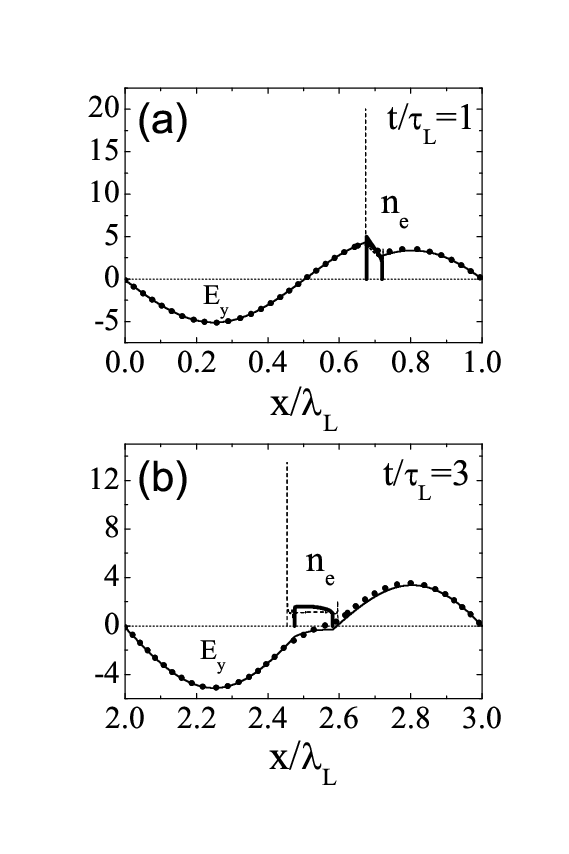}}
\caption{Density profiles of electron layer \protect$n(x)$ (thick
lines) inserted into electric field $E_{Ly}(x)$ of single-cycle
laser pulse (thin lines) at times (a) one and (b) three laser cycles
after first interaction with foil. Model results (solid) are
compared with 1D-PIC results (dashed). Density is normalized to
critical density and electric field to \protect$E_0=mc\omega_L/e$
(see text).} \label{fig1}
\end{figure}

The layer of finite thickness is divided into 100 sub-layers,
and each sub-layer is assumed to move according to Eqs.~(\ref{py-eq}-\ref{x-eq}).
Finally, the density is calculated, using $x(t,x_0)$ and
\begin{equation}
n = n_0/(\partial x/\partial x_0). \label{den}
\end{equation}
From this we obtain spatial density distributions, as shown in
Fig.~\ref{fig1} and ~\ref{fig3}.

\begin{figure}[b]
\centering\resizebox{1\columnwidth}{!}{\includegraphics{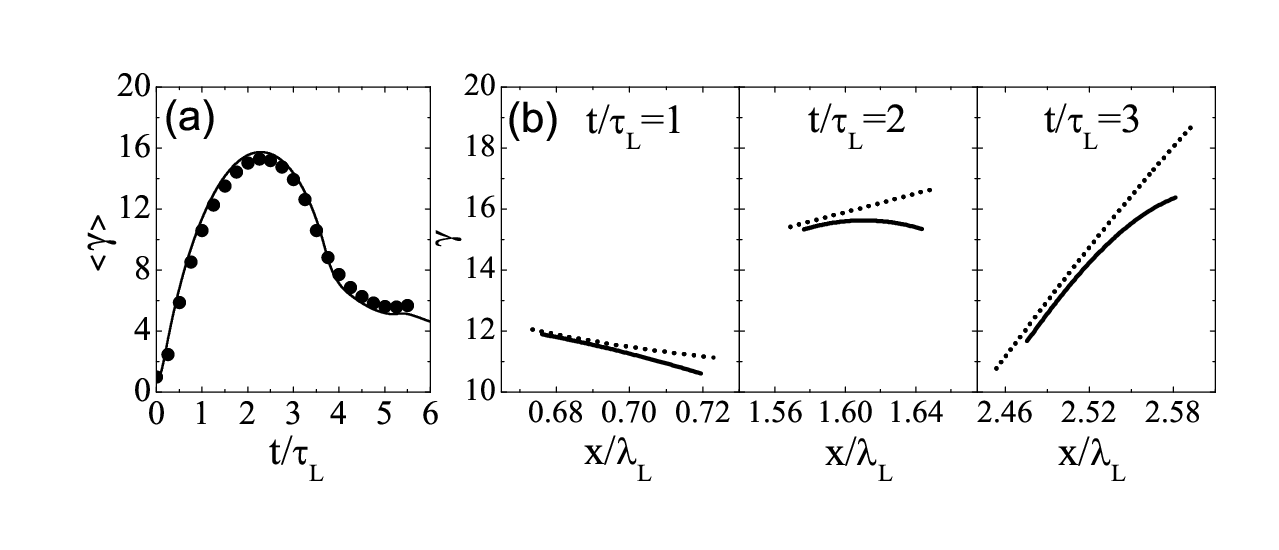}}
\caption{(a) Temporal evolution of the $<\gamma>$-factor averaged
over electron layer. (b) Spatial distribution of $\gamma$-factor
over electron layer at different times. Model results (solid lines)
are compared with 1D-PIC results (dashed lines).}
\label{fig2}
\end{figure}

\section{Comparison of model with PIC simulation}

The model is now applied to the reference case, already discussed in paper I \cite{paper I}.
It is compared with 1D-PIC simulation in Figs.~\ref{fig1} and \ref{fig2}.
A planar foil with density $N=n_0/n_{crit}=159$ and $\varepsilon_0=Nd_0=1$
is irradiated by a single-cycle laser pulse $E_{Ly}=a_0\sin\tau$ ($0\le \tau\le 2\pi$)
with $a_0=5$. At this high intensity, we consider foil ionization to happen very rapidly;
actually we use the approximation of an initially completely ionized plasma layer.
The 1D-PIC simulations have been performed using the code LPIC++ \cite{lpic}.

The results in Fig.~\ref{fig1} show profiles of laser field and electron density
at times $t/\tau_0=1$ and $t/\tau_0=3$ after interaction with the foil.
The laser pulse has expelled all electrons from the foil.
They are seen as a dense layer carried along by the first half-wave of the pulse,
while foil ions are considered immobile and are located at $x=0$ (not shown).
Apparently, the electron layer is transparent to the light, the front has penetrated
the layer, but is depleted due to interaction with the layer.
Different from the results in paper I, the analytical model \cite{Kulagin}
is now well reproducing the 1D-PIC results. There are two reasons for this agreement:
(1) correct account for the self-radiation of the electron layer
which may be viewed as the effect of forward Thomson scattering
reducing the wave amplitude in front of the layer;
(2) correct account of initial electron acceleration
inside the ion layer.

A conspicuous deviation between model and PIC results concerns the density spike
seen in the PIC results on the side of the incident light pulse.
It appears in simulations with different PIC codes and may be of numerical origin.
This needs further study.
Smaller deviations in the layer thickness develop at later times
(see Fig. 1b at $t/\tau_L=3$) and may be attributed to approximations
made in Eqs. (\ref{Esy}) and (\ref{Bsz}) for the self-radiation fields.

In Figs. 2a and 2b, the temporal evolution of the electron $\gamma$-factor
is depicted. Again we find good agreement between model and PIC results,
both for values $\langle \gamma \rangle$ averaged over the layer (Fig. 2a) and
spatial $\gamma$ gradients (Fig. 2b).  The average rises up to
a maximum of $\langle \gamma \rangle \approx 15$,
and then falls again when the layer drifts into the decelerating phase
of the second half-wave. Notice that the maximum of $\langle \gamma \rangle$
is much lower than the pure single-particle estimate
of $\gamma_{max}=2a_0^2=50$. This is because of the electrostatic and
electromagnetic self-fields of the layer, well described by the model.

\section{Layer compression depending on initial thickness}

\begin{figure}[tbp]
\centering\resizebox{1\columnwidth}{!}{\includegraphics{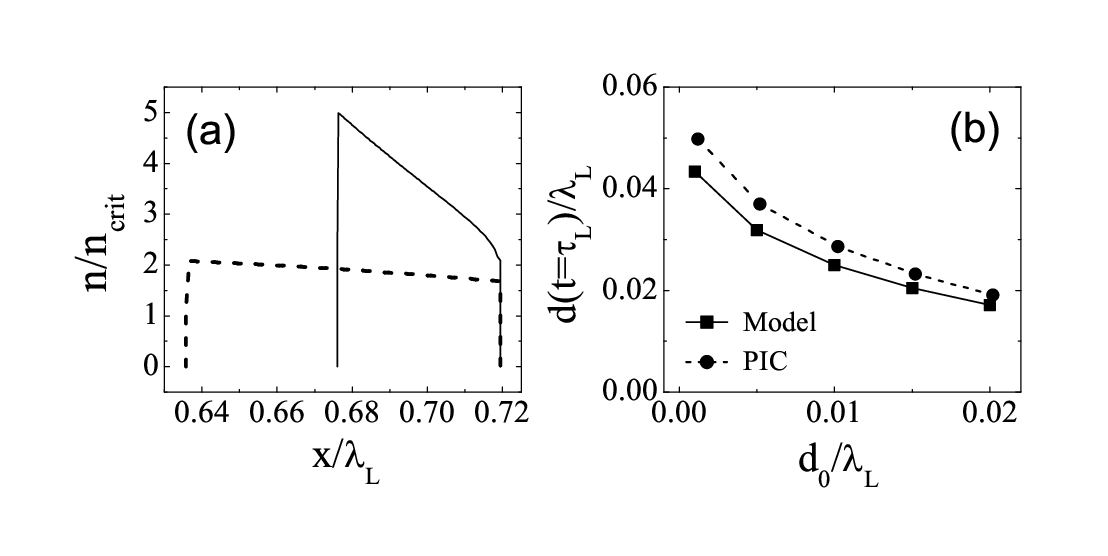}}
\caption{(a) Density distribution at $t/\tau_L=1$ with (solid line)
and without (dashed line) accounting for the linear increase
of $E_{sx}$ inside the ion volume (compare Eq.~(\ref{Esx})).
(b) Thickness of electron layer $d$ at time $t/\tau_L=1$
plotted versus initial thickness $d_0$; model results
(solid line) compared to 1D-PIC results (dashed line).}
\label{fig3}
\end{figure}

A particularly interesting observation is a tendency for layer compression
during the first stage of acceleration stage.
Compression may even overcome Coulomb expansion.
It originates from the initial phase of acceleration,
when the electrons are still inside the ion layer. There the $E_x$ field,
felt by an electron initially at $x_0$ and given by Eq.~(\ref{Esx}),
rises linearly $\propto (x-x_0)$ before reaching a constant $E_x$ value
when leaving the ion volume. During this initial phase,
electrons on the left side of the layer facing the incident laser pulse
gain more energy than those to the right. This leads to layer bunching,
as it is demonstrated for time $t/\tau_L=1$ in Fig.~\ref{fig3}a.
Here the dashed profile corresponds to using $E_{sx}(x,x_0)=N(d_0-x_0)$
for all $x$ and $x_0$, while the solid profile corresponds
to the correct treatment using full Eq.~(\ref{Esx}).

The compression depends on the initial thickness of the foil.
It is more pronounced when starting with thicker foils.
In Fig.~\ref{fig3}b, we show the thickness $d$ at time $t/\tau_L=1$ as a function
of initial thickness $d_0$, keeping the areal density $Nd_0=1$ constant.
For $d_0/\lambda_L=0.02$ and $t/\tau_L=1$, layer expansion is not only
reduced relative to the reference case with $d_0/\lambda_L=0.001$, but is
actually slightly compressed. This effect is seen in both model and PIC results.
This is good news for experiments using laser pulses with more realistic shapes.
In case of Gaussian shapes rather than the sharp-front flat-top pulses considered here,
one expects some extent of foil expansion before the pulse maximum hits.
The compression then leads to high-density layers during the first laser cycles of
layer acceleration.

\section{Expansion of a freely propagating layer}

\begin{figure}[t]
\centering\resizebox{1.0\columnwidth}{!}{\includegraphics{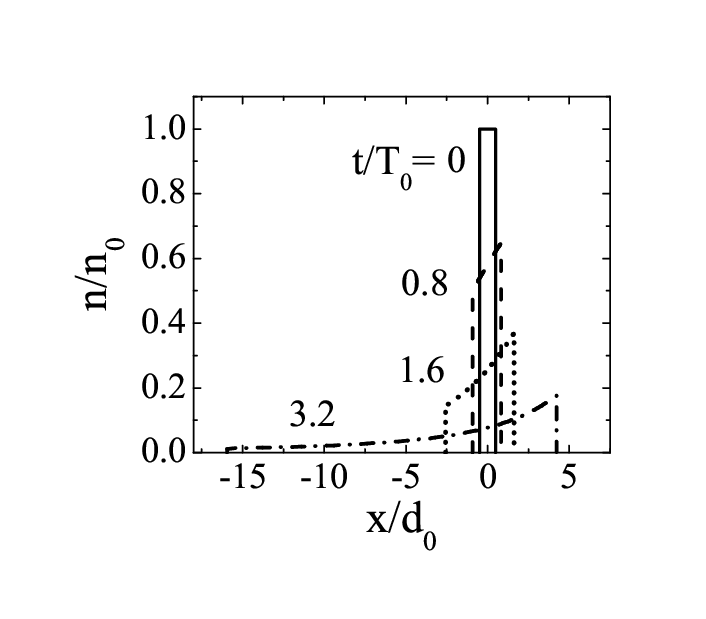}}
\caption{Longitudinal Coulomb expansion of electron layer with
initial density $n_0= 2.7\times10^{23}$ cm$^{-3}$ and thickness $d_0
= 0.8$ nm, moving at $\gamma_0  = 10$. Density profiles are plotted
for times $t/T_0=0$ (solid), 0.8 (dashed), 1.6 (dotted), and 3.2
(dash-dotted). The characteristic decay time is $T_0=4.3$ fs.}
\label{fig4}
\end{figure}

Finally we discuss a simple solution which is important for the late stage
of a freely propagation layer when it is not in contact any more
with the laser field. The solution describes the longitudinal decay
of the relativistic mirror due to Coulomb explosion.
Initially the layer has uniform density $n_0$ and plasma frequency $\omega_p$.
It consists of electrons only (no ion layer) and propagates in $x$-direction
with momentum $\gamma_0\beta_0$. The equation of motion \ref{eom} then reduces to
\begin{equation}
dp_x/dt=Nx_0
\end{equation}
keeping definitions and normalization the same as before.
The coordinate $x_0$ now denotes the initial position of a test electron
in the plane layer extending from $-d_0$ to $+d_0$.
The longitudinal momentum is
\begin{equation}
p_x(t,x_0)=\beta_0\gamma_0 + Nx_0 t,
\end{equation}
and the corresponding energy
\begin{equation}
\gamma=\gamma_0\sqrt{1 + 2\beta_0\frac{x_0}{d_0}\frac{t}{T_0}
+\left(\frac{x_0}{d_0}\frac{t}{T_0}\right)^2}
\end{equation}
involves a characteristic time $T_0=\gamma_0/(Nd_0)$.
In dimensional units, it is given by
\begin{equation}
T_0=\frac{\gamma_0}{\omega_p^2d_0/c}.
\end{equation}
Apparently, it is the relativistic plasma frequency
which sets the longitudinal decay time.
The electron trajectories $x(t,x_0)$ can be easily inferred from energy
conservation $\gamma-\gamma_0=Nd_0(x-x_0)$, giving
\begin{equation}
x(t,x_0)=x_0+(\gamma-\gamma_0)/(Nd_0).
\end{equation}
From this the evolution of the density profile $n(x,t)$ is obtained
with the help of Eq.~(\ref{den}). Profiles of an initially uniform
electron layer are plotted in Fig.~\ref{fig4}.
In this case the characteristic decay time is $T_0=4.3$ fs.

\section{Conclusion}

In conclusion, we have considered electron acceleration
from ultra-thin foils by high-contrast ultra-short laser pulses
in the regime in which all electrons are expelled from the foil.
They form a dense electron layer that can be used
as a relativistic mirror for Thomson backscattering, at least
over a short time interval of a few laser periods after foil interaction.
This paper supplements a companion paper discussing the backscattered
spectra \cite{paper I}. Here we have shown that the analytical model
of Kulagin et al. \cite{Kulagin} well reproduces corresponding
particle-in-cell simulations. Analytical theory describing
layer dynamics is important for basic understanding as well as
developing and analyzing future experiments.

As a particular result, we have identified
a regime of layer compression related to initial acceleration
inside the ion volume.
It is found that foils somewhat expanded initially are
superior to foils of same areal density, but thinner and denser.
Also an analytical formula is given for longitudinal layer
expansion after it disconnects from the driving laser field
and is freely propagating.

\section{Acknowledgments}
Meng Wen thanks for the opportunity to work for one year at
Max-Planck-Institute for Quantum Optics and acknowledges support
from the Max-Planck-Gesellschaft and the Chinese Academy of
Sciences. This work was also supported by the DFG project Transregio
TR18, by the Munich Centre for Advanced Photonics (MAP), and by the
Association EURATOM - Max-Planck-Institute for Plasma Physics.

\end{document}